\documentclass[12pt]{article}
\usepackage[dvips]{graphicx}
\usepackage{epsfig,amsmath}
\begin{document}
\thispagestyle{empty}
\begin{center}

{\bf NEW DEVELOPMENTS IN THE QUANTUM STATISTICAL APPROACH OF THE PARTON DISTRIBUTIONS \footnote{Invited talk presented at the Workshop DUBNA-SPIN 09, Sept. 01-05, 2009, Dubna (Russia), to appear in the Proceedings.}}

\vskip1.40cm
{\bf Jacques Soffer}
\vskip 0.3cm
{\it Physics Department, Temple University\\
Barton Hall, 1900 N, 13th Street\\
Philadelphia, PA 19122-6082, USA} 
\vskip 1.0cm
{\bf Abstract}
\end{center}

We briefly recall the main physical features of the parton distributions in the quantum statistical picture of the nucleon.
Some predictions from a next-to-leading order QCD analysis are compared to recent experimental results.  

\vspace{7.2mm} 
 A new set of parton distribution functions (PDF) was constructed in the framework of a statistical approach of the nucleon \cite{bbs1}, which has the following characteristic features:
\begin{itemize}
\item For quarks (antiquarks), the building blocks are the helicity dependent distributions $q_{\pm}(x)$ ($\bar q_{\pm}(x)$) and we define $q(x)= q_{+}(x)+q_{-}(x)$ and $\Delta q(x) = q_{+}(x)-q_{-}(x)$ (similarly for antiquarks).
\item At the initial energy scale taken at $Q^2_0= 4 \mbox{GeV}^2$, these distributions are given
by the sum of two terms, a quasi Fermi-Dirac function and a helicity independent diffractive
contribution, which leads to a universal behavior at very low $x$ for all flavors.
\item The flavor asymmetry for the light sea, {\it i.e.} $\bar d (x) > \bar u (x)$, observed in the data
is built in. This is clearly understood in terms of the Pauli exclusion principle, based on the fact that the proton contains two $u$ quarks and only one $d$ quark.
\item The chiral properties of QCD lead to strong relations between $q(x)$ and $\bar q (x)$.
For example, it is found that the well estalished result $\Delta u (x)>0 $\ implies $\Delta 
\bar u (x)>0$ and similarly $\Delta d (x)<0$ leads to $\Delta \bar d (x)<0$.
\item Concerning the gluon, the unpolarized distribution $G(x,Q_0^2)$ is given in terms of a quasi
 Bose-Einstein function, with only {\it one free parameter}, and for simplicity, one assumes zero gluon polarization, {\it i.e.} $\Delta G(x,Q_0^2)=0$, at the initial energy scale $Q_0^2$.
\item All unpolarized and polarized  light quark distributions depend upon {\it eight}
free parameters, which were determined in 2002 (see  Ref.~\cite{bbs1}), from an NLO fit of a selected set of accurate DIS data.
\end{itemize}
More recently, new tests against experimental (unpolarized and polarized)
data turned out to be very satisfactory, in particular in hadronic reactions, as reported in Refs.~\cite{bbs2,bbs3,bbs-next}.
The statistical approach has been extended \cite{bbs4} to the interesting sitution where the PDF have, in addition to the usual Bjorken $x$ dependence, an explicit $k_T$ transverse momentum dependence (TMD) and this might be used in future calculations with no $k_T$ integration. This 
will not be treated here for lack of time, although it is a very important topic, with a growing interest because it is now clear that several new phenomena are sensitive to TMD effects.

\begin{center}
\begin{figure}[h]
\vspace{-1pc}
\begin{minipage}{18pc}
\vspace{-4pc}
\hspace{-2pc}
\includegraphics[width=20pc]{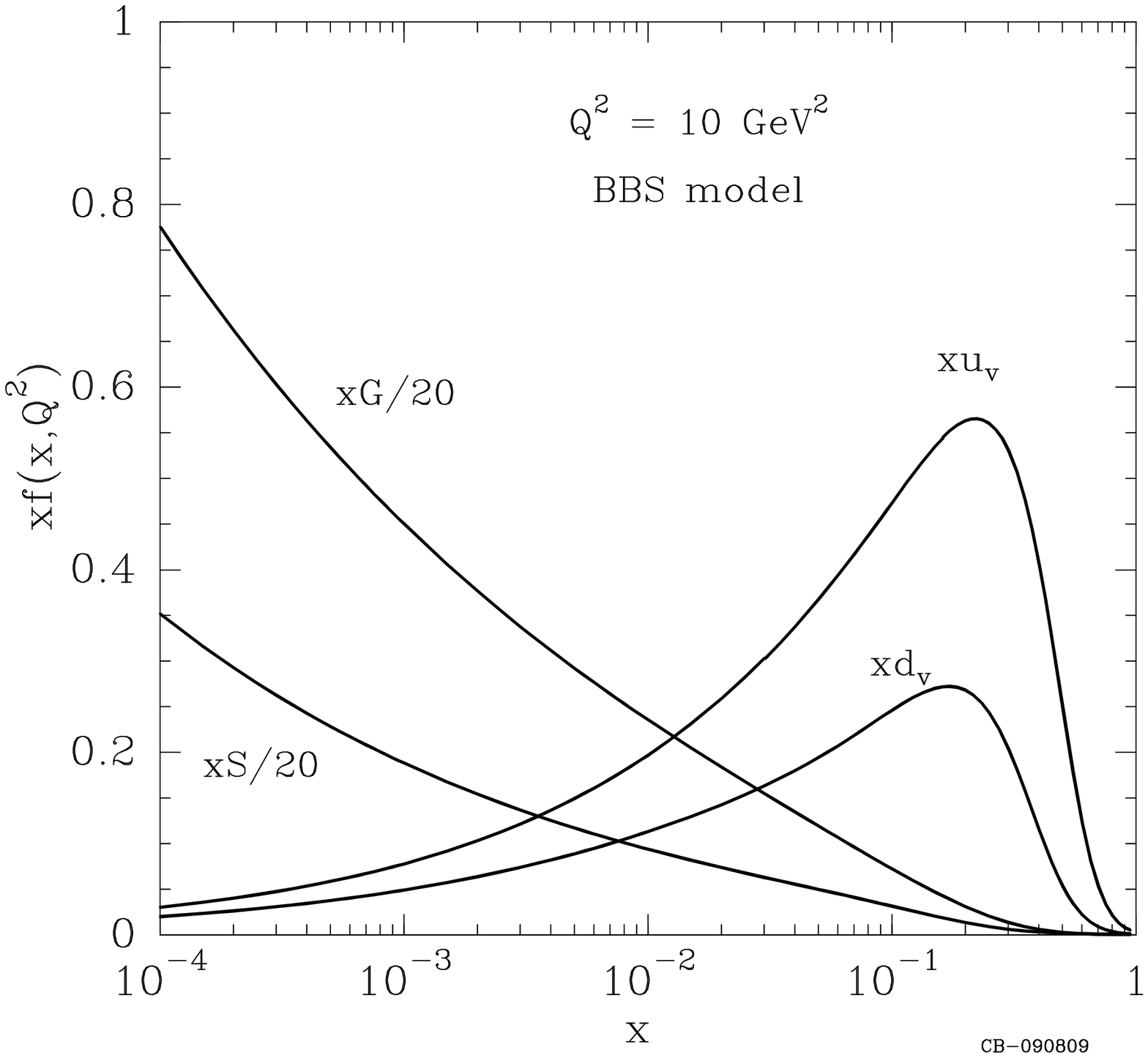}
\caption{\label{label}
BBS predictions for various statistical parton
distributions versus $x$, at $Q^2=10 \mbox{GeV}^2$.}
\end{minipage}
\hspace{2pc}
\begin{minipage}{18pc}
\includegraphics[width=19pc]{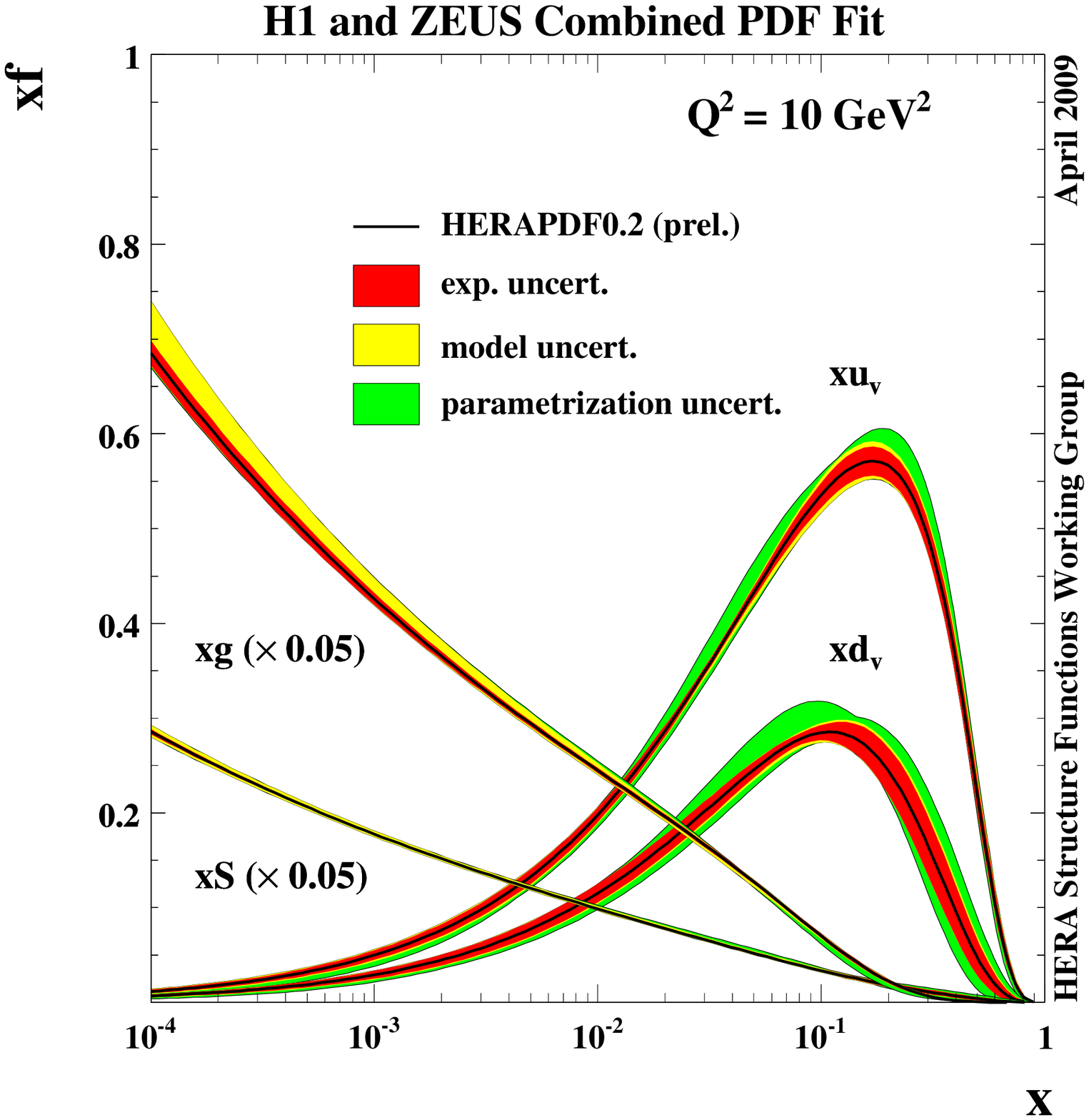}
\caption{\label{label}
Parton distributions at $Q^2$=10 $\mbox{GeV}^2$ as determined by the H1PDF fit, with different
uncertainties ( Taken from Ref.~\cite{pdf-h1}).}
\label{fi:pdf}
\end{minipage} 
\end{figure}
\end{center}
We display on Fig.~1 the resulting unpolarized statistical PDF versus $x$ at $Q^2$=10 $\mbox{GeV}^2$, where $xu_v$ is the $u$-quark valence,
$xd_v$ the $d$-quark valence, $xG$ the gluon and $xS$
stands for twice the total antiquark contributions, $\it i.e.$ $xS(x)=2x(\bar {u}(x)+ \bar {d}(x) + \bar {s}(x))+ \bar {c}(x))$. Note that $xG$ and $xS$ are downscaled by a factor 0.05. They can be compared with the parton distributions as determined by the H1PDF 2009 QCD NLO fit, shown in Fig.~2, and the agreement is rather good. The results are based on recent $ep$ collider data combined with previously published data and the accuracy is typically
in the range of 1.3 - 2 $\%$.\\

 In the statistical approach the unpolarized gluon distribution has a very simple expression given by
$xG(x,Q_0^2)=\frac{A_Gx^{b_G}}{\exp(x/{\bar x})-1}$, where $\bar x=0.099$ is the universal temperature, $A_G=20.53$ is determined by the momentum sum rule and $b_G=0.90$ is the only free parameter. It is consistent with the available data, mostly coming indirectly from the QCD $Q^2$ evolution of $F_2(x,Q^2)$, in particular in the low $x$ region. However it is
known that $ep$ DIS cross section is characterised by two independent structure funtions, $F_2(x,Q^2)$ and the longitudinal structure function $F_L(x,Q^2)$. For low
$Q^2$, the contribution of the later to the cross section at HERA is only sizeable at $x$ smaller than approximately $10^{-3}$ and in this domain the gluon density dominates over the sea quark density. More precisely, it was shown that using some approximations \cite{amcs} one has, 
$xG(x,Q^2) = \frac{3}{10}5.9[\frac{3\pi}{2\alpha_s}F_L(0.4x,Q^2)-F_2(0.8x,Q^2)] \simeq \frac{8.3}{\alpha_s}F_L(0.4x,Q^2)$. Before
HERA was shut down, a dedicated run period, with reduced proton beam energy, was
approved, allowing H1 to collect new results on $F_L$. We show on Fig.~3 the expectations
of the statistical approach compared to the new data, whose precision is reasonable. The trend and the magnitude of the prediction are in fair agreement with the data, so this is another test of the good predictive power of our theoretical framework.\\
\begin{center}
\begin{figure}[h]
\begin{minipage}{18pc}
\hspace{-3pc}
\includegraphics[angle=-90,width=20pc]{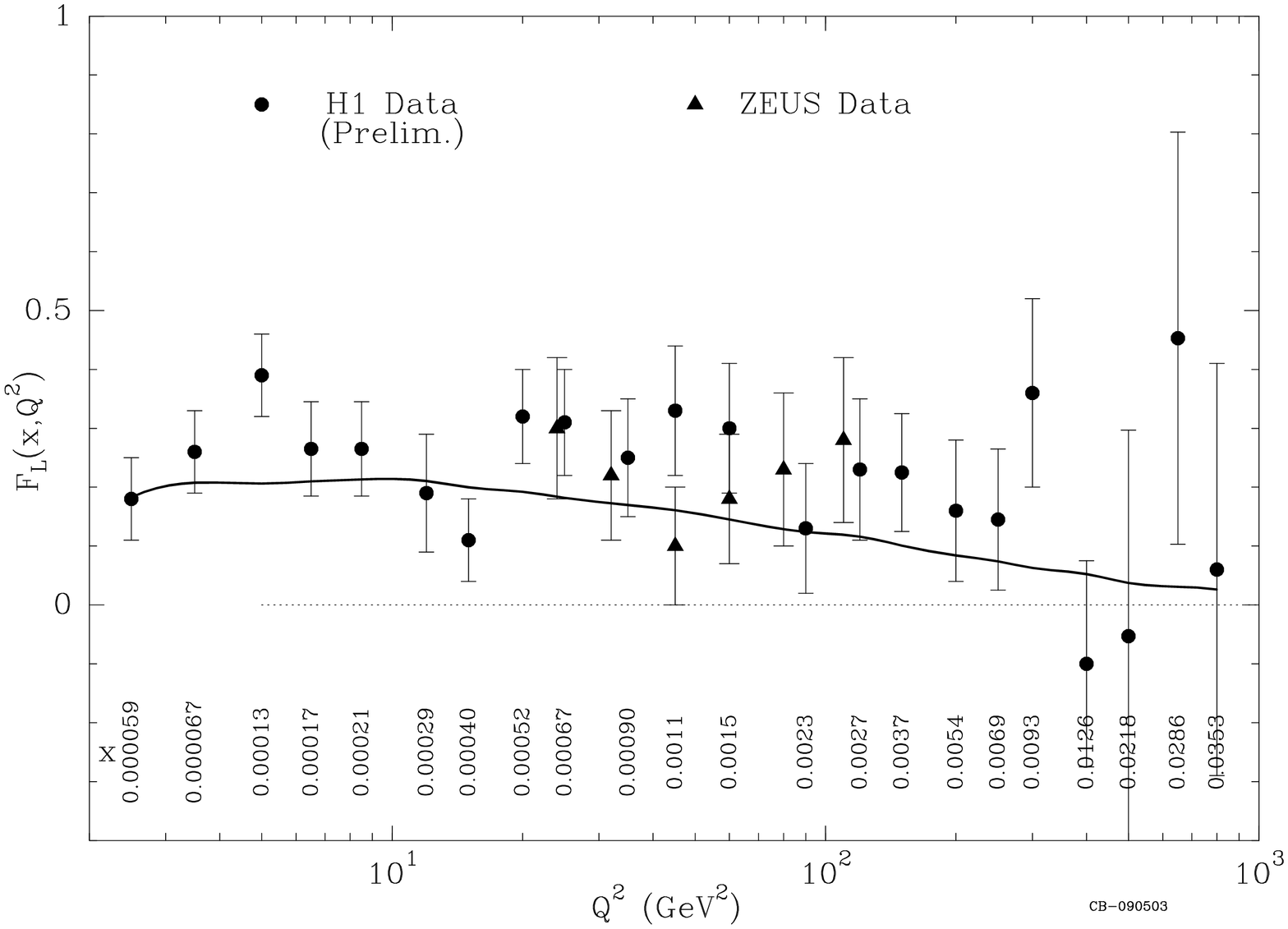}
\caption{\label{label}
The longitudinal proton structure function $F_L(x,Q^2)$ averaged in $x$ at given values of $Q^2$. Data from \cite{fl-h1,fl-zeus} compared to
the BBS theoretical prediction.}
\end{minipage}\hspace{2pc}%
\begin{minipage}{18pc}
\hspace{-2pc}
\includegraphics[angle=-90,width=21pc]{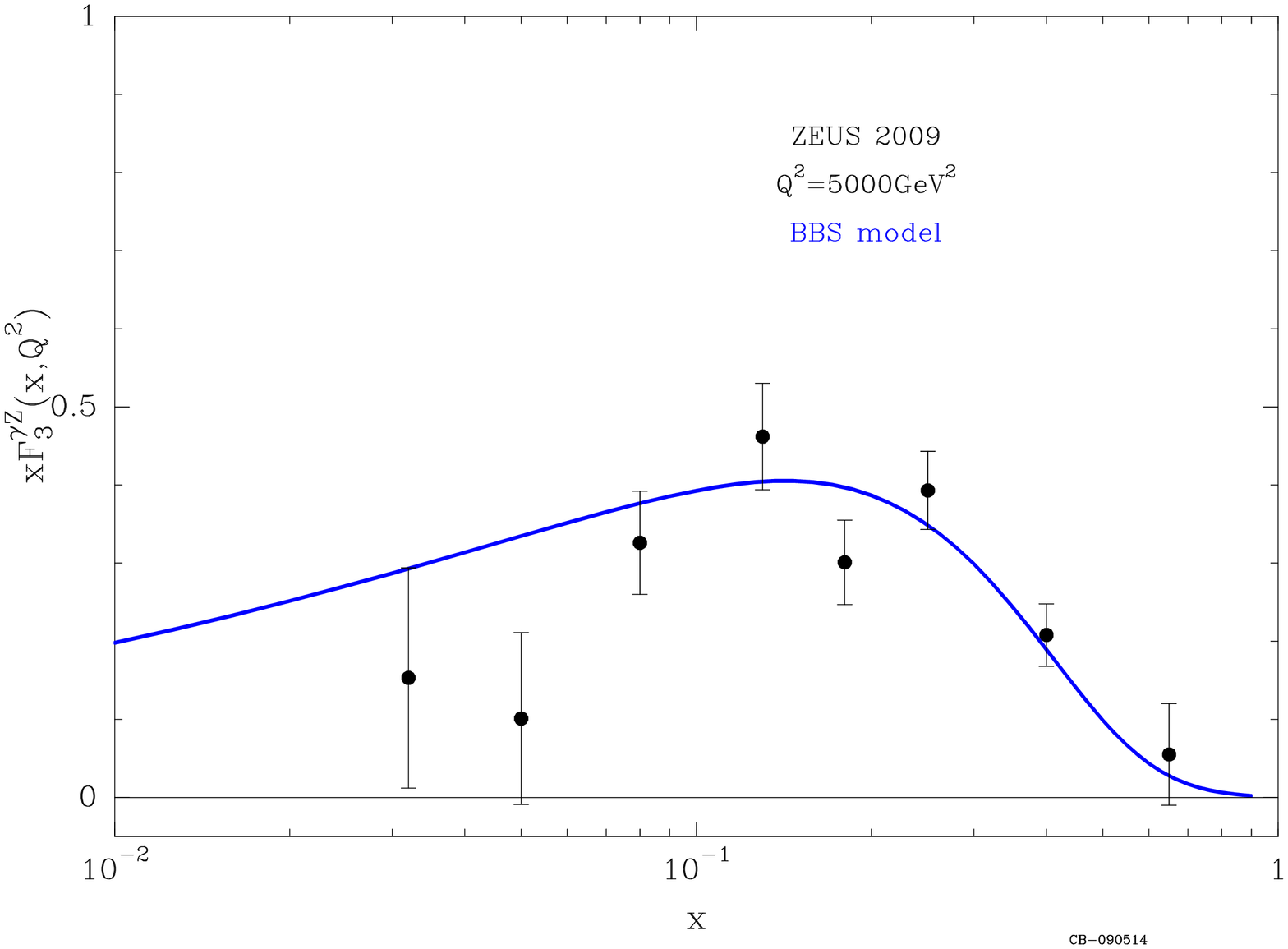}
\caption{\label{label}
The interference term $xF_3^{\gamma Z}$ extracted in $e^{\pm}p$ collisions at HERA. Data from \cite{fgz-zeus} compared to the BBS prediction.}
\label{fi:inter}
\end{minipage} 
\end{figure}
\end{center}

\begin{center}
\begin{figure}[h]
\begin{minipage}{18pc}
\vspace{-3pc}
\includegraphics[width=15pc]{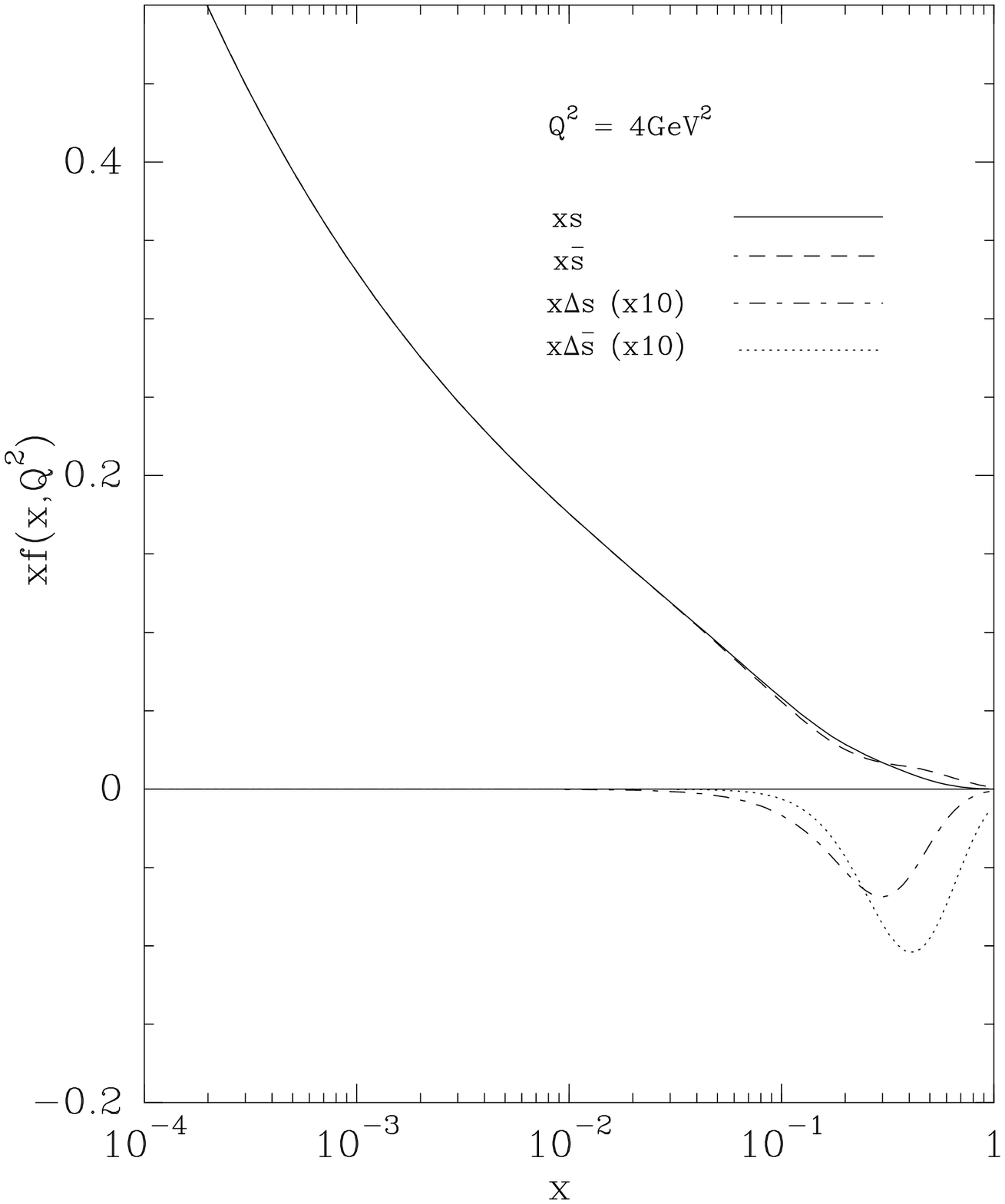}
\caption{\label{label}
Strange quark and antiquark distributions determined at NLO.}
\end{minipage}
\hspace{+1pc}
\begin{minipage}{18pc}
\hspace{-1pc}
\includegraphics[angle=-90,width=20pc]{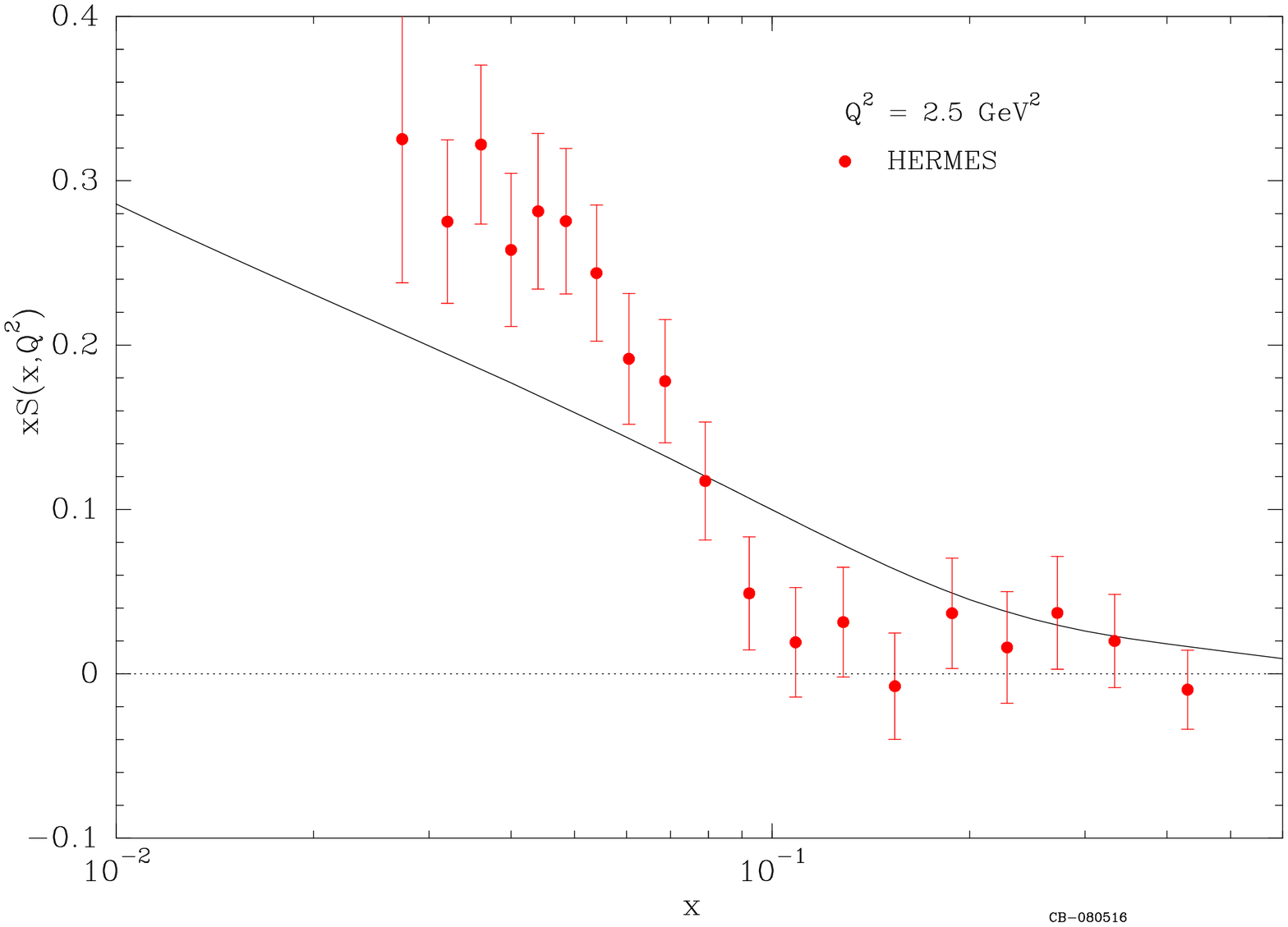}
\caption{\label{label}
The strange parton distribution $xS(x)=xs(x)+x{\bar s}(x)$ at $Q^2$=2.5 $\mbox{GeV}^2$. The theoretical prediction from \cite{bbs5} is compared to data from \cite{hermes}.}
\label{fi:strange}
\end{minipage} 
\end{figure}
\end{center}
One can also test the behavior of the interference term between the photon and the $Z$ exchanges, which can be isolated in neutral current $e^{\pm}p$ collisions at high $Q^2$. We have to a good approximation, if sea quarks are ignored, $xF_3^{\gamma Z}= \frac{x}{3}(2u_v + d_v)$ and the comparison between data and prediction is displayed in Fig.~4.\\
Concerning the strange quark and antiquark distributions, a careful analysis of the NuTeV CCFR data led us to the conclusion that $s(x,Q^2)\neq \bar {s}(x,Q^2)$ and the corresponding polarized distributions are unequal, small and negative \cite{bbs5}, as shown in Fig.~5. The rapid rise one observes
in the small $x$ region is compatible with the data from Hermes, as shown in Fig.~6.\\
Finally let us recall that the subject of quark and antiquark transversity distribution in the proton is also a very interesting topic. By studying a possible connection
between helicity and transversity, we have proposed a simple toy model (see Ref.~\cite{bbs6}).\\

{\bf Acknowledgments}\\
I am grateful to the organizers of DSPIN09 for their warm hospitality at JINR and for their invitation to present this talk. My special thanks go to Prof. A.V. Efremov for providing a full financial support and for making, once more, this meeting so successful.


\begin{thebibliography}{99} 
\bibitem{bbs1}
C.~Bourrely, F.~Buccella, J.~Soffer, Euro. Phys. J. {\bf C23} (2002)~487.\\
For a practical use of these PDF, we refer the reader to the following web site:\\
www.cpt.univ-mrs.fr/${\sim}$ bourrely/research/bbs-dir/bbs.html.
 
\bibitem{bbs2}
C.~Bourrely, F.~Buccella, J.~Soffer, Mod. Phys. Letters {\bf A18} (2003)~771. 
 
\bibitem{bbs3}
C.~Bourrely, F.~Buccella, J.~Soffer, Euro. Phys. J. {\bf C41} (2005)~327. 

\bibitem{bbs-next}
C.~Bourrely, F.~Buccella, J.~Soffer, (in preparation). 

\bibitem{bbs4}
C.~Bourrely, F.~Buccella, J.~Soffer, Mod. Phys. Letters {\bf A21} (2006)~143.

\bibitem{pdf-h1}
F.D. Aaron, {\it et al.}, H1 Collaboration, DESY-09-005, arXiv:0904.3513 [hep-ex],\\ to appear EPJC. 

\bibitem{amcs} 
A. M. Cooper-Sarkar et {\it al.}, Z. Phys. {\bf C39}, (1988)~281.

\bibitem{fl-h1}
H1 Collaboration, submitted to DIS 2009, Madrid, Spain, April 26-30 (2009).

\bibitem{fl-zeus}
S. Chekanov, {\it et al.}, ZEUS Collaboration, Phys. Lett. {\bf B682} (2009)~8.

\bibitem{fgz-zeus}
S. Chekanov, {\it et al.}, ZEUS Collaboration, Euro. Phys. J. {\bf C62} (2009)~625.
 
\bibitem{bbs5}
C.~Bourrely, F.~Buccella, J.~Soffer, Phys. Lett. {\bf B648} (2007)~39.

\bibitem{hermes}
A.~Airapetian, {\it et al.}, Hermes Collaboration, Phys. Lett. {\bf B666} (2008)~446.

\bibitem{bbs6}
C.~Bourrely, F.~Buccella, J.~Soffer, Mod. Phys. Letters {\bf A24} (2009)~1889.


\end{thebibliography}
\end{document}